\documentclass[english,aps,manuscript]{revtex4}

\usepackage[T1]{fontenc}
\usepackage[latin9]{inputenc}
\usepackage{amsmath}
\usepackage{amssymb}
\usepackage{graphicx}
\usepackage{esint}

\makeatletter
\@ifundefined{textcolor}{}
{%
 \definecolor{BLACK}{gray}{0}
 \definecolor{WHITE}{gray}{1}
 \definecolor{RED}{rgb}{1,0,0}
 \definecolor{GREEN}{rgb}{0,1,0}
 \definecolor{BLUE}{rgb}{0,0,1}
 \definecolor{CYAN}{cmyk}{1,0,0,0}
 \definecolor{MAGENTA}{cmyk}{0,1,0,0}
 \definecolor{YELLOW}{cmyk}{0,0,1,0}
}

\makeatother

\usepackage{babel}
\begin{document}

\title{Initial states and infrared physics in locally de Sitter spacetime}

\author{Klaus Larjo and David A. Lowe}

\email{klaus_larjo,lowe@brown.edu}

\address{Department of Physics, Box 1843, Brown University, Providence, RI,
02912, USA}
\begin{abstract}
The long wavelength physics in a de Sitter region depends on the initial
quantum state. While such long wavelength physics is under control
for massive fields near the Hartle-Hawking vacuum state, such initial
states make unnatural assumptions about initial data outside the region
of causal contact of a local observer. We argue that a reasonable
approximation to a maximum entropy state, one that makes minimal assumptions
outside an observer's horizon volume, is one where a cutoff is placed
on a surface bounded by timelike geodesics, just outside the horizon.
For sufficiently early times, such a cutoff induces secular logarithmic
divergences with the expansion of the region. For massive fields,
these effects sum to finite corrections at sufficiently late times.
The difference between the cutoff correlators and Hartle-Hawking correlators
provides a measure of the theoretical uncertainty due to lack of knowledge
of the initial state in causally disconnected regions. These differences
are negligible for primordial inflation, but can become significant
during epochs with very long-lived de Sitter regions, such as we may
be entering now.
\end{abstract}
\maketitle

\section{Introduction}

At the classical level, de Sitter spacetime appears to be stable.
For conformally-coupled matter rigorous nonlinear stability theorems
have been established that show for an open set of initial data the
solution evolves to an asymptotically de Sitter solution at late times
\cite{FRIEDRICH:1986fk,Friedrich:1986kx,Friedrich:1991nn}.

Quantum mechanically, the situation is less clear. In a quantum theory
correlators do depend on data outside the past light cone of points.
It is therefore important to study the effect of choice of initial
state on predictions for cosmological correlators, that then in turn
determine the spectrum of density perturbations at late times.

There has been much confusion in the literature on this point. Broadly
speaking there are two main camps, those who advocate using a dS invariant
quantum state, to compute dS invariant correlators; and those who
impose more physically motivated initial conditions, breaking the
de Sitter symmetries. In the first scenario, the slow roll parameters
provide the only means of explicit breaking of the de Sitter isometries.
It has been convincingly argued that at least in massive scalar theories,
these correlators are infrared finite %
\footnote{For conformally coupled matter, this was studied long ago \cite{Adler:1972qq,Adler:1973ty},
and for more general matter in \cite{Drummond:1975yc,Drummond:1977dg,Drummond:1977uy}.
These questions have been revisited more recently in \cite{Marolf:2010nz,Marolf:2010zp,Hollands:2010pr,Marolf:2011sh,Hollands:2011we}.%
}. It remains an open question whether the same is true when the quantum
fluctuations of gravitons are included.

In the other camp are those who have typically had in mind direct
applications to inflation. In that case, an initial state is chosen
at some finite time, breaking the de Sitter invariance. Needless to
say, the correlators break de Sitter invariance, and in massless scalar
theories, exhibit infrared divergences. Infrared cutoffs lead to terms
in the correlators that grow as $\log a(t)$ to some power \cite{Traschen:1986tn,Tsamis:1992sx,Brandenberger:1993zc,Tsamis:1994ca,Tsamis:1996qm,Tsamis:1996qq,Mukhanov:1996ak,Abramo:1997hu,Abramo:1998hi,Weinberg:2005vy,Weinberg:2006ac,Polyakov:2006bz,vanderMeulen:2007ah,Bartolo:2007ti,Polyakov:2007mm,Adshead:2009cb,Burgess:2009bs,Polyakov:2009nq,Krotov:2010ma}
(here $a(t)$ is the scale factor in the Friedmann-Robertson-Walker
metric). In this setting there have also been computations involving
gravitons at one-loop, which also find the $\log a(t)$ growth \cite{Tsamis:1996qk}.
This has led to the suggestion that quantum effects may lead to a
relaxation of the cosmological constant \cite{Tsamis:1992sx}. See
\cite{Seery:2010kh} for a review article summarizing many of these
approaches. Some also \cite{Kahya:2009sz,Kundu:2011sg} for some observations
closely related to the approach discussed in the present paper.

In this paper we will argue the main results of these two camps are
mutually compatible, and that the difference between the two approaches
may be viewed as a theoretical uncertainty in the relevant correlator.
Since it is likely we do not have causal access to the spacetime region
that would allow us to precisely determine the initial state, the
best we can do is perform some version of a Bayesian analysis and
quantify the theoretical uncertainty in the initial state, and hence
the derived correlators.

If we work within the general framework of inflation, the initial
state with the minimal number of assumptions that gives rise to our
present observable universe, is a state that started as an approximately
homogeneous region an inverse $10^{14}$ GeV in size (see \cite{Kolb:1990vq}
for a more detailed discussion of the initial conditions for inflation).
To explain the horizon problem, this region must inflate to a size
of about 1 m, assuming inflation ends, and reheating produces a thermal
gas with temperature around $10^{14}$ GeV (this gives the famous
60 e-foldings of inflation needed for viability). Subsequently the
universe evolves according to the Standard Model of cosmology. Thus,
at least classically, we can conclude that any state that respects
these conditions will give rise to what we see today. According to
the Bayesian approach, we should pick the most likely such state (or
maximum entropy state). The most obvious such state would involve
the homogeneous patch, but would be surrounded by a thermal gas at
the Planck temperature. Of course such a state would involve large
gravitational back-reaction, and it would be impossible to compute
with it using known methods %
\footnote{The landscape of string theory provides a wealth of viable vacua typically
separated by Planck scale potential walls. This fits well with the
picture of eternal inflation, where a Planck scale curvature foam
tends to dominate the volume on any given space like slice. It is
worth emphasizing that approximate approaches to performing computations
in these situations, such as stochastic inflation \cite{Starobinsky:1986fx,Starobinsky:1994bd},
where super horizon fluctuations in a fixed de Sitter background are
considered in order to compute an initial wavefunction for the universe,
or Coleman-de Luccia bubble nucleation \cite{Coleman:1980aw}, where
tunneling to an open universe sets the initial conditions for inflation,
both require extensive fine tuning. In \cite{Starobinsky:1986fx,Starobinsky:1994bd},
because only the dynamics of superhorizon modes is considered, the
high curvature spacetime foam is ignored, and replaced by a fixed
de Sitter background, which is an inconsistent approximation. Likewise
if we wish to apply the results of \cite{Coleman:1980aw} to eternal
inflation in the string landscape we must assume we are in a special
state where the semiclassical instanton methods are applicable. The
philosophy in the present paper is to sidestep these issues, by making
minimal assumptions about the initial state of inflation, in order
to better model this generic strongly curved foam outside the initial
homogeneous region. %
}. In the following we will adopt a compromise that leaves one with
computable correlators, but removes most assumptions about the initial
region outside the homogeneous region. Therefore we propose to use
an initial state with a hard infrared cutoff just outside the horizon
at the start of inflation when the homogeneous patch begins to expand.

After this time, the homogeneous patch expands. If the infrared cutoff
was kept at a fixed proper length, the patch would soon exceed the
size of the cutoff, and the resulting correlators would miss much
of the physics of interest to us today. Moreover, to keep the cutoff
at the same length, the walls of the box would have to contract faster
than the speed of light, which signals an unphysical choice of regulator.
Nevertheless, such a cutoff is needed for massless theories in pure
dS spacetime to retain the dS invariance of the correlators.

The more physical choice we advocate in the present work is to view
the infrared cutoff as a kind of bubble wall. The trajectory of the
wall will depend on the details of the bubble, but for any physical
choice, at late times, the wall will asymptote to a family of timelike
geodesics. Therefore the simplest cutoff that captures these qualitative
features is a comoving cutoff. Such a cutoff has been studied before
in the literature in \cite{Lyth:2007jh,Bartolo:2007ti}, where it
is referred to as a minimal box. Thus we advocate using a comoving
infrared cutoff placed just outside the horizon at the beginning of
inflation. This explicitly breaks de Sitter invariance. As we will
see, this can lead to important effects at late times in correlators
of massless fields. 

The dS invariant computations, on the other hand, represent the maximal
number of assumptions about the initial state. Choosing the Euclidean
vacuum across the entire inflating patch amounts to solving the horizon
problem by hand by imposing gaussian fluctuations of fixed size across
the entire initial slice. The chief advantage of doing this is that,
at least for massive scalar theories (where no additional infrared
cutoff is needed), the correlators are dS invariant, and can be much
more simply treated to extract predictions for observations today.
Moreover, as we shall see, the answers agree with the leading terms
of the correlators with a comoving cutoff, provided inflation does
not last too long. The difference between the two approaches should
be interpreted as a measure of the theoretical uncertainty in the
correlator due to our ignorance of the initial state at the start
of inflation.

As we shall see, the $\log a(t)$ terms in the examples we study are
subleading versus the dS invariant results provided inflation does
not last extraordinarily long. Thus predictions for primordial inflation
are largely unchanged. 

On the other hand, we may be entering a regime of dS dominance in
our present epoch. Assuming the graviton is the only field that experiences
these $\log a(t)$ divergences, we can expect this quantum instability
of de Sitter space to become relevant only after a very large number
of e-foldings. We note this instability is predicted from a unitary
model of quantum gravity based on embedding dS regions in asymptotically
AdS spacetimes, dual to conformal field theories \cite{Lowe:2007ek,Lowe:2010np}.

\section{SCALAR FIELD THEORY USING THE IN-IN FORMULATION}

To illustrate the interpretation of amplitudes mentioned above, we
take examples from $\phi^{4}$ scalar field theory. Our goal is to
exhibit the one-loop corrections computed using the comoving infrared
cutoff (our approximate maximum entropy initial state) and compare
them to computations done without an infrared cutoff, choosing the
usual Euclidean vacuum state. It is also necessary to specify an ultraviolet
cutoff to regulate the one-loop integrals. We choose an ultraviolet
cutoff motivated by local effective field theory, simply a cutoff
at fixed proper momentum in both cases. While these or closely related
results have already appeared in the literature \cite{vanderMeulen:2007ah,Enqvist:2008kt,Burgess:2009bs},
we review the derivation in some detail to establish a consistent
notation and collect all the relevant results together.

We work with the Lagrangian 
\begin{align}
\mathcal{L} & =\sqrt{-g}\left(-\frac{1}{2}\partial_{\mu}\phi\partial^{\mu}\phi-\frac{1}{2}m^{2}\phi^{2}-\frac{\lambda}{4!}\phi^{4}\right)+\delta\mathcal{L\qquad\mathrm{with}}\label{eq:Lagrangian}\\
 & \delta\mathcal{L}=\sqrt{-g}\left(-\frac{1}{2}\delta_{Z}\partial_{\mu}\phi\partial^{\mu}\phi-\frac{1}{2}\delta_{m}\phi^{2}-\frac{\delta_{\lambda}}{4!}\phi^{4}\right),\nonumber 
\end{align}
where $\delta\mathcal{L}$ contains the counter terms. The de Sitter
metric is written as 
\[
ds^{2}=-dt^{2}+a(t)^{2}d\vec{x}^{2}=a(\tau)^{2}(-d\tau^{2}+d\vec{x}^{2}),
\]
with $a(t)=e^{Ht},$ and conformal time is defined as $\tau=-H^{-1}a(t)^{-1}$.
We use the notation $x=(\vec{x},t)$.

\subsection{Mode expansions}

To set up the perturbative expansion, we begin with the expansion
of the free field in modes labeled by comoving wavevectors $k$ 
\begin{align}
 & \phi^{(0)}(\vec{x},\tau)=\int\frac{d^{3}\vec{k}}{(2\pi)^{3}}\left[e^{i\vec{k}\cdot\vec{x}}\phi_{k}(\tau)\alpha_{\vec{k}}+e^{-i\vec{k}\cdot\vec{x}}\phi_{k}^{*}(\tau)\alpha_{\vec{k}}^{\dag}\right],\quad{\rm with}\label{k-field}\\
 & \phi_{k}(\tau)=-\frac{\sqrt{-\pi\tau}}{2a(\tau)}H_{\nu}^{(1)}(-k\tau).\label{mode}
\end{align}
Here $H_{\nu}^{(1)}$ is the Hankel function of the first kind and
$\nu^{2}\equiv9/4-m^{2}/H^{2}$. The $\alpha_{\vec{k}}$ satisfy standard
canonical commutation relations. The Bunch--Davies vacuum is defined
by $\alpha_{\vec{k}}|0\rangle=0$, for all $\vec{k}$. %
\footnote{It should be noted that a more general family of de Sitter invariant
vacua \cite{Allen:1985ux}, known as the alpha vacua, are possible
for quantum fields on a fixed de Sitter background. See for example
\cite{Goldstein:2002fc,Goldstein:2003qf,Goldstein:2003ut} and references
therein. However this approach is likely to break down when quantum
gravity fluctuations are included. This is essentially because the
Euclidean/Bunch-Davies vacuum is the only one that preserves the desired
short distance singularities of correlators, whereas the alpha vacuum
correlators contain additional singularities at antipodal points,
creating light-like non-locality. This in turn can lead to strong
back-reaction effects once spacetime fluctuations are included. %
}

\subsection{In-In formalism}

\begin{figure}
\begin{centering}
\includegraphics{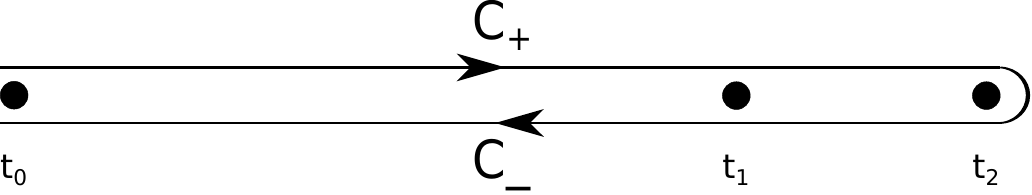}
\par\end{centering}

\caption{This shows the Keldesh time-ordering contour $C_{0}=C_{+}\cup C_{-}$,
for the case $t_{2}>t_{1}$. The initial state is specified at time
$t_{0}$. \label{fig:Keldesh}}
\end{figure}

We will work with the in-in formalism, where the goal is to compute
expectation values of time-ordered products of Heisenberg field operators
with respect to our fixed initial state. This differs from the usual
path integral approach, which computes transition amplitudes. A nice
review of the in-in, or Schwinger-Keldesh approach can be found in
\cite{Maciejko:2007vn}. The perturbative expansion can be derived
using interaction picture methods as usual 
\[
\langle0|T\left(\phi(\vec{x}_{1},t_{1})\phi(\vec{x}_{2},t_{2})\right)|0\rangle=\langle0|T_{C_{0}}\left(e^{-i\int_{t_{0}}^{\infty}H_{{\rm int}}(t')dt'}\phi^{(0)}(\vec{x}_{1},t_{1})\phi^{(0)}(\vec{x}_{2},t_{2})\right)|0\rangle
\]
with the new feature being the appearance of the Schwinger-Keldesh
time-ordering operator $T_{C_{0}}$ which accomplishes the task of
time evolving the amplitude from the initial time $t_{0}$, to $\max(t_{1},t_{2})$
and then back again to $t_{0}$ so that the expectation value is computed.
The contour $C_{0}$ is shown in figure \ref{fig:Keldesh}. The operator
insertions at $t_{1}$ and $t_{2}$ appear on the upper contour $C_{+}$
. Here $H_{{\rm int}}$ is the interacting part of the Hamiltonian
obtained from \eqref{eq:Lagrangian}.

When the exponential is expanded in powers of $\lambda$ we can apply
Wick's theorem to break the expectation value into integrals of products
of the following elementary time-ordered two-point functions: 
\[
G(x_{1},x_{2})=\left\{ \begin{array}{ll}
G^{T}(x_{1},x_{2})\equiv-i\langle T[\phi(\vec{x}_{1},t_{1})\phi(\vec{x}_{2},t_{2})]\rangle, & t_{1},t_{2}\in C_{+},\\
G^{<}(x_{1},x_{2})\equiv\phantom{-}i\langle\phi(\vec{x}_{2},t_{2})\phi(\vec{x}_{1},t_{1})\rangle, & t_{1}\in C_{+},\,\,\, t_{2}\in C_{-},\\
G^{>}(x_{1},x_{2})\equiv-i\langle\phi(\vec{x}_{1},t_{1})\phi(\vec{x}_{2},t_{2})\rangle, & t_{1}\in C_{-},\,\,\, t_{2}\in C_{+},\\
G^{\bar{T}}(x_{1},x_{2})\equiv-i\langle\bar{T}[\phi(\vec{x}_{1},t_{1})\phi(\vec{x}_{2},t_{2})]\rangle, & t_{1},t_{2}\in C_{-}.
\end{array}\right.
\]
We then have the relations 
\begin{align*}
 & G^{T}(x_{1},x_{2})=\Theta(t_{1}-t_{2})G^{<}(x_{1},x_{2})+\Theta(t_{2}-t_{1})G^{>}(x_{1},x_{2}),\\
 & G^{\bar{T}}(x_{1},x_{2})=\Theta(t_{1}-t_{2})G^{>}(x_{1},x_{2})+\Theta(t_{2}-t_{1})G^{<}(x_{1},x_{2}).
\end{align*}
The contour $C_{0}$ can be written as a single $\mathbb{R}$ by writing
\[
\phi(\vec{x},t)=\left\{ \begin{array}{ll}
\phi_{+}(\vec{x},t), & t\in C_{+},\\
\phi_{-}(\vec{x},t), & t\in C_{-},
\end{array}\right.
\]
and the two-point functions can be combined into a 2$\times$2 matrix
${\bf G}$ as 
\[
{\bf G}=\left(\begin{array}{cc}
G^{T} & G^{<}\\
G^{>} & G^{\bar{T}}
\end{array}\right)=\left(\begin{array}{cc}
-i\langle T[\phi_{+}(\vec{x}_{1},t_{1})\phi_{+}(\vec{x}_{2},t_{2})]\rangle & i\langle\phi_{+}(\vec{x}_{1},t_{1})\phi_{-}(\vec{x}_{2},t_{2})\rangle\\
i\langle\phi_{-}(\vec{x}_{1},t_{1})\phi_{+}(\vec{x}_{2},t_{2})\rangle & -i\langle\bar{T}[\phi_{-}(\vec{x}_{1},t_{1})\phi_{-}(\vec{x}_{2},t_{2})]\rangle
\end{array}\right).
\]
It is convenient to work in a different basis in field space by defining
\[
\left(\begin{array}{c}
\phi_{C}\\
\phi_{\Delta}
\end{array}\right)=\left(\begin{array}{c}
\frac{1}{2}(\phi_{+}+\phi_{-})\\
\phi_{+}-\phi_{-}
\end{array}\right)={\bf R}\left(\begin{array}{c}
\phi_{+}\\
\phi_{-}
\end{array}\right),\quad{\rm with}\quad{\bf R}=\left(\begin{array}{cc}
\frac{1}{2} & \frac{1}{2}\\
1 & -1
\end{array}\right),
\]
and in this basis the matrix of correlators becomes 
\begin{align*}
 & {\bf G_{K}}={\bf RGR}^{T}=\left(\begin{array}{cc}
iF & G^{R}\\
G^{A} & 0
\end{array}\right),\quad{\rm with}\\
 & iF(x_{1},x_{2})\equiv\frac{1}{2}\left(G^{>}(x_{1},x_{2})+G^{<}(x_{1},x_{2})\right),\\
 & G^{R}(x_{1},x_{2})\equiv\Theta(t_{1}-t_{2})\left(G^{<}(x_{1},x_{2})-G^{>}(x_{1},x_{2})\right),\\
 & G^{A}(x_{1},x_{2})\equiv\Theta(t_{2}-t_{1})\left(G^{>}(x_{1},x_{2})-G^{<}(x_{1},x_{2})\right).
\end{align*}

\subsection{The Feynman rules and the one-loop diagrams}

\begin{figure}

\begin{centering}
\includegraphics[scale=0.6]{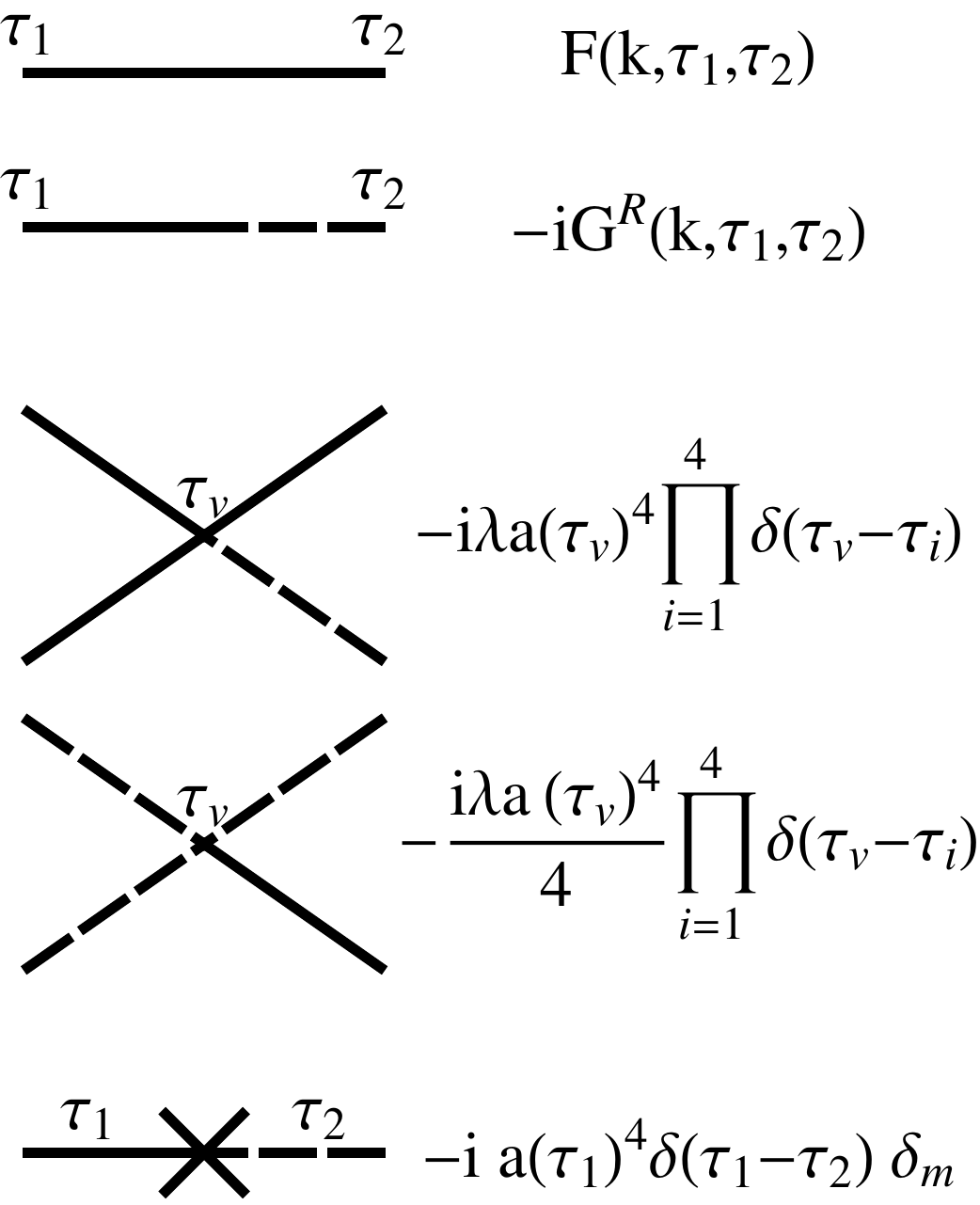}
\par\end{centering}

\caption{The Feynman rules.\label{fig-feyn}}

\end{figure}

In the $(\phi_{C},\phi_{\Delta})$-basis the Lagrangian becomes 
\begin{align*}
\mathcal{L} & =\mathcal{L}(\phi_{+})-\mathcal{L}(\phi_{-})=\sqrt{-g}\left(-\partial_{\mu}\phi_{C}\partial^{\mu}\phi_{\Delta}-m^{2}\phi_{C}\phi_{\Delta}-\frac{\lambda}{4!}(\phi_{C}\phi_{\Delta}^{3}+4\phi_{C}^{3}\phi_{\Delta})\right)\\
 & +\sqrt{-g}\left(-\delta_{Z}\partial_{\mu}\phi_{C}\partial^{\mu}\phi_{\Delta}-\delta_{m}\phi_{C}\phi_{\Delta}-\frac{\delta_{\lambda}}{4!}(\phi_{C}\phi_{\Delta}^{3}+4\phi_{C}^{3}\phi_{\Delta})\right)
\end{align*}
and we read off the Feynman rules of figure \ref{fig-feyn}.

\begin{figure}[t]
\begin{centering}
\includegraphics[scale=0.55]{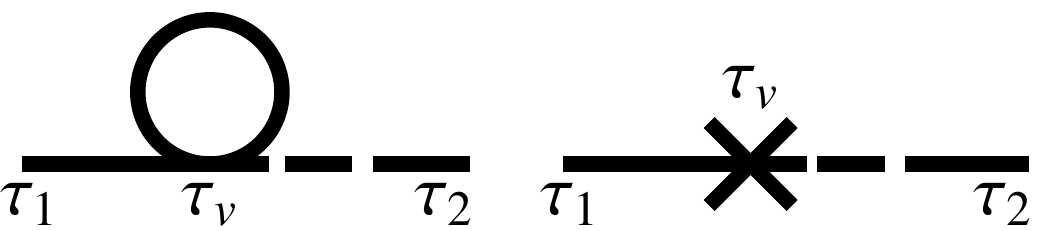} 
\par\end{centering}

\centering{}\caption{The diagrams contributing to $F(x_{1},x_{2})$ at order $\lambda.$\label{fig-loop} }
\end{figure}

The Feynman rules related to the counter-terms $\delta_{Z}$ and $\delta_{\lambda}$
yield additional powers of the scale factor $a(t)$ and coupling $\lambda$
respectively, and will not contribute to a late time, one-loop computation.
Hence we only include the Feynman rule corresponding to $\delta_{m}.$
Since the equal time propagator $G^{R}(x,x)$ vanishes by construction,
any non-vanishing one-loop diagram has to have a solid line inside
the loop. Thus by the above Feynman rules the only contributions at
one-loop level are given in figure \ref{fig-loop}. According to the
Feynman rules, these diagrams contribute

\begin{align}
F_{1L}(x_{1},x_{2}) & =\int\frac{d^{3}\vec{k}}{(2\pi)^{3}}e^{i\vec{k}\cdot(\vec{x}_{1}-\vec{x}_{2})}F_{1L}(k,\tau_{1},\tau_{2}),\quad{\rm with}\label{x-space}\\
F_{1L}(k,\tau_{1},\tau_{2}) & =\int d\tau_{v}F(k,\tau_{1},\tau_{v})\left(-iG^{R}(k,\tau_{2},\tau_{v})\right)\left(\mathrm{L}(\tau_{v})+\mathrm{C}(\tau_{v})\right),\label{k-space}\\
{\rm L}(\tau_{v}) & =-i\lambda a(\tau_{v})^{4}\int\frac{d^{3}\vec{p}}{(2\pi)^{3}}F(p,\tau_{v},\tau_{v}),\quad C(\tau_{v})=-ia(\tau_{v})^{4}\delta_{m},\nonumber 
\end{align}
where $\mathrm{L}(\tau_{v})$ is the contribution from the amputated
one-loop diagram, and $\mathrm{C}(\tau_{v})$ comes from the counter-term
diagram.

\subsection{The loop contribution}

As explained in the introduction, we consider a comoving infrared
cut-off $\Lambda_{{\rm IR}}$ and a physical ultraviolet cutoff $\Lambda_{{\rm UV}}$.
The contribution from the loop integral can then be written as 
\begin{equation}
\mathrm{L}(\tau_{v})=-i\lambda a(\tau_{v})^{4}\left(\int_{\Lambda_{IR}}^{-\tau_{v}^{-1}}+\int_{-\tau_{v}^{-1}}^{\Lambda_{UV}a(\tau_{v})}\right)\frac{dp}{(2\pi)^{2}}2p^{2}F(p,\tau_{v},\tau_{v}),\label{loop1}
\end{equation}
where we split the integral into parts corresponding to modes inside
and outside the horizon at time $\tau_{v}$. In these regions the
propagators can be expanded using the mode functions (\ref{mode})
as 
\begin{align}
\textrm{Outside: }|k\tau_{i}|\ll1 & \quad\left\{ \begin{array}{l}
F(k,\tau_{1},\tau_{2})\approx\frac{H^{2}}{2k^{3}}(k^{2}\tau_{1}\tau_{2})^{\epsilon},\\
G^{R}(k,\tau_{1},\tau_{2})\approx\Theta(\tau_{1}-\tau_{2})\frac{H^{2}}{3}\left[\tau_{1}^{3-\epsilon}\tau_{2}^{\epsilon}-\tau_{2}^{3-\epsilon}\tau_{1}^{\epsilon}\right],
\end{array}\right.\label{smallk}\\
\textrm{Inside: }|k\tau_{i}|\gg1 & \quad\left\{ \begin{array}{l}
F(k,\tau_{1},\tau_{2})\approx\frac{H^{2}\tau_{1}\tau_{2}}{2k}\cos k(\tau_{1}-\tau_{2}),\\
G^{R}(k,\tau_{1},\tau_{2})\approx\Theta(\tau_{1}-\tau_{2})\frac{H^{2}\tau_{1}\tau_{2}}{k}\sin k(\tau_{1}-\tau_{2}),
\end{array}\right.\label{largek}
\end{align}
where $\epsilon\equiv m^{2}/(3H^{2})$ is the mass parameter. Inserting
the expansions into the loop integral (\ref{loop1}) we get 
\begin{align}
{\rm L}(\tau_{v}) & \approx\frac{-i\lambda}{(2\pi)^{2}H^{2}\tau_{v}^{4}}\left(\tau_{v}^{2\epsilon}\int_{\Lambda_{IR}}^{-\tau_{v}^{-1}}dp\,\, p^{-1+2\epsilon}+\tau_{v}^{2}\int_{-\tau_{v}^{-1}}^{\Lambda_{UV}a(\tau_{v})}dp\,\, p\right)\nonumber \\
 & \approx\frac{-i\lambda}{(2\pi)^{2}H^{2}\tau_{v}^{4}}\left(\frac{1-(\Lambda_{{\rm IR}}\tau_{v})^{2\epsilon}}{2\epsilon}+\frac{1}{2}\left(\frac{\Lambda_{{\rm UV}}}{H}\right)^{2}\right).\label{loop2}
\end{align}
 Note that the integrals are dominated by the IR and UV regions as
opposed to the horizon $|p\tau_{v}|\sim1$, and the approximations
(\ref{smallk},\ref{largek}) can be used. In order to cancel the
ultraviolet divergence we fix the counter-term coefficient $\delta_{m}$
to be

\[
\delta_{m}=-\frac{\lambda H^{2}}{2(2\pi)^{2}}\left[\left(\frac{\Lambda_{UV}}{H}\right)^{2}-\left(\frac{\mu}{H}\right)^{2}\right],
\]
where $\mu$ is the renormalization scale. This leads to the contribution

\begin{equation}
\mathrm{L}(\tau_{v})+\mathrm{C}(\tau_{v})=\frac{-i\lambda}{(2\pi)^{2}H^{2}\tau_{v}^{4}}\left(\frac{1-(\Lambda_{{\rm IR}}\tau_{v})^{2\epsilon}}{2\epsilon}+\frac{1}{2}\left(\frac{\mu}{H}\right)^{2}\right)\equiv\frac{-i\lambda}{(2\pi)^{2}H^{2}\tau_{v}^{4}}\mathrm{V}(\tau_{v}).\label{eq:loop3}
\end{equation}

\subsection{The propagator at late times}

\begin{figure}[t]
\centering{}\includegraphics[scale=0.7]{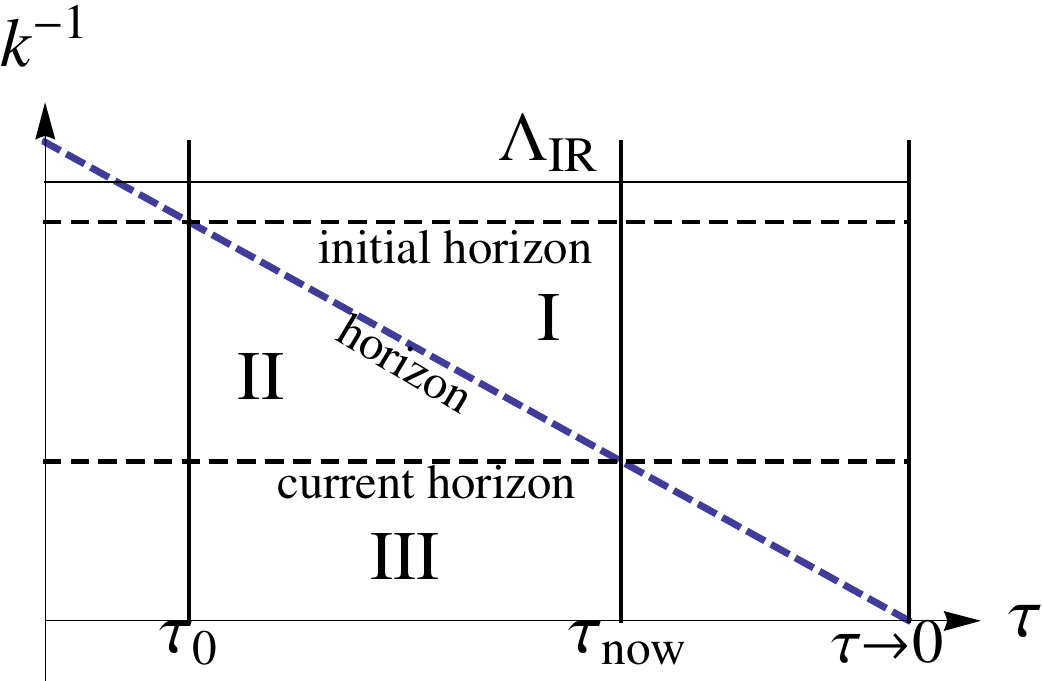} \caption{Plot of the integration region in $(\tau,k)$-spac\label{fig-reg}e.}
\end{figure}

We then wish to analyze the late time dependence of the propagator
\eqref{x-space}. Inserting the loop contribution \eqref{eq:loop3}
we get 
\begin{equation}
F_{1L}(x_{1},x_{2})=\frac{-\lambda}{(2\pi)^{2}H^{2}}\int\frac{d^{3}\vec{k}}{(2\pi)^{3}}e^{i\vec{k}\cdot(\vec{x}_{1}-\vec{x}_{2})}\int\frac{d\tau_{v}}{\tau_{v}^{4}}F(k,\tau_{1},\tau_{v})G^{R}(k,\tau_{2},\tau_{v})\mathrm{V}(\tau_{v}).\label{eq:prop-basic}
\end{equation}
The integration region naturally splits into components depending
on whether mode $k$ is inside or outside the horizon, as indicated
in figure \ref{fig-reg}. In the following we will take $\tau_{1}<\tau_{2}$
for convenience, and also assume that they are in the same cosmological
period, $\tau_{1}\sim\tau_{2}\sim\tau_{\mathrm{now}}.$ The limit
$\tau_{2}\gg\tau_{1}$ is also of interest, and we will treat it separately
in the following.

\paragraph{The IR modes:}

Region I consists of modes outside the horizon. Using the expansions
(\ref{smallk}) we find the momentum space propagator to be 
\[
F_{1L}^{I}(k,\tau_{1},\tau_{2})=-\frac{\lambda H^{2}}{(2\pi)^{2}}\int_{\tau_{X}}^{\tau_{2}}\frac{d\tau_{v}}{\tau_{v}^{4}}\frac{(k^{2}\tau_{1}\tau_{v})^{\epsilon}}{2k^{3}}\frac{1}{3}\left[\tau_{2}^{3-\epsilon}\tau_{v}^{\epsilon}-\tau_{v}^{3-\epsilon}\tau_{2}^{\epsilon}\right]\mathrm{V}(\tau_{v}),
\]
where the lower limit of integration is given by $\tau_{X}\equiv{\rm Max}(\tau_{0},-k^{-1})$,
see figure \ref{fig-reg}. This can be evaluated analytically, yielding
\begin{align}
 & F_{1L}^{I}(k,\tau_{1},\tau_{2})=-\frac{\lambda H^{2}}{(2\pi)^{2}}\frac{(k^{2}\tau_{1}\tau_{2})^{\epsilon}}{6k^{3}}\left\{ \frac{(\Lambda_{IR}\tau_{2})^{2\epsilon}}{2\epsilon}\left[\frac{1-\left(\frac{\tau_{2}}{\tau_{X}}\right)^{3-4\epsilon}}{3-4\epsilon}+\frac{1-\left(\frac{\tau_{X}}{\tau_{2}}\right)^{2\epsilon}}{2\epsilon}\right]\right.\nonumber \\
 & \left.-\left(\frac{1}{2}\left(\frac{\mu}{H}\right)^{2}+\frac{1}{2\epsilon}\right)\left(\frac{1-\left(\frac{\tau_{2}}{\tau_{X}}\right)^{3-2\epsilon}}{3-2\epsilon}-\log\frac{\tau_{X}}{\tau_{2}}\right)\right\} \equiv-\frac{\lambda H^{2}}{(2\pi)^{2}}\frac{(k^{2}\tau_{1}\tau_{2})^{\epsilon}}{6k^{3}}D(\tau_{X}).\label{F-kspace}
\end{align}
Transforming back to position space we have 
\begin{align*}
F_{1L}^{I}(x_{1},x_{2}) & =\int\frac{d^{3}\vec{k}}{(2\pi)^{3}}e^{i\vec{k}\cdot(\vec{x}_{1}-\vec{x}_{2})}F_{1L}^{I}(k,\tau_{1},\tau_{2})=\frac{2}{(2\pi)^{2}}\int_{\Lambda_{IR}}^{-\tau_{2}^{-1}}dk\frac{\sin(k\Delta x)}{k\Delta x}k^{2}F_{1L}^{I}(k,\tau_{1},\tau_{2})\\
 & \approx\frac{2}{(2\pi)^{2}}\int_{\Lambda_{IR}}^{-\tau_{2}^{-1}}dkk^{2}F_{1L}^{I}(k,\tau_{1},\tau_{2})(1+\mathcal{O}((k\Delta x)^{2}),
\end{align*}
where $\Delta x\equiv|\vec{x}_{1}-\vec{x}_{2}|,$ and in the last
step we took the physical separation between $\vec{x}_{1}$ and $\vec{x}_{2}$
to be much less than the Hubble distance. Inserting (\ref{F-kspace})
we find 
\begin{align}
F_{1L}^{I}(x_{1},x_{2}) & =-\frac{\lambda H^{2}(\tau_{1}\tau_{2})^{\epsilon}}{(2\pi)^{4}}\int_{\Lambda_{{\rm IR}}}^{-\tau_{2}^{-1}}dk\frac{k^{2\epsilon}}{3k}D(\max(\tau_{0},-k^{-1}))\nonumber \\
 & =-\frac{\lambda H^{2}(\tau_{1}\tau_{2})^{\epsilon}}{3(2\pi)^{4}}\left(D(\tau_{0})\int_{\Lambda_{{\rm IR}}}^{-\tau_{0}^{-1}}dkk^{-1+2\epsilon}+\int_{-\tau_{0}^{-1}}^{-\tau_{2}^{-1}}dkk^{-1+2\epsilon}D(-k^{-1})\right).\label{eq:Fir}
\end{align}
 The first integral is over modes outside the horizon already at $\tau_{0}$,
and vanishes if one takes the cut-off to coincide with the initial
horizon. The integrals can be computed analytically to yield for late
times
\begin{align}
F_{1L}^{I}(x_{1},x_{2}) & =-\frac{\lambda H^{2}}{12(2\pi)^{4}\epsilon}\left\{ \left(\frac{\tau_{1}\tau_{2}}{\tau_{0}^{2}}\right)^{\epsilon}\log\frac{\tau_{0}}{\tau_{2}}\left[\left(\frac{\mu}{H}\right)^{2}+\frac{1}{\epsilon}\right]\left(1-(\Lambda_{{\rm IR}}\tau_{0})^{2\epsilon}\right)\right.\nonumber \\
 & \left.+\frac{(\Lambda_{{\rm IR}}^{2}\tau_{1}\tau_{2})^{\epsilon}}{\epsilon}\log\frac{\tau_{0}}{\tau_{2}}+\frac{1}{2\epsilon}\left(\frac{\tau_{1}}{\tau_{2}}\right)^{\epsilon}\left[\left(\frac{\mu}{H}\right)^{2}+\frac{1}{\epsilon}\right]\left(1-\left(\frac{\tau_{2}}{\tau_{0}}\right)^{2\epsilon}\right)\right\} .\label{eq:late-ir}
\end{align}
The first two terms vanish for late times, while the last one asymptotes
to a constant value that is independent of the infrared cutoff $\Lambda_{{\rm IR}}$. 

The last term in \eqref{eq:late-ir} looks worrisome, as it is divergent
in the limit where one takes the leg $\tau_{2}$ to future infinity
while keeping the other leg $\tau_{1}$ fixed. This is due to our
assumption that the legs are in the same cosmological period, which
no longer holds when $\tau_{2}$ is taken to the far-future. The divergence
arises because the horizons at $\tau_{2}$ and $\tau_{1}$ no longer
coincide in this limit, but rather modes exit the $\tau_{1}$ horizon
earlier than the $\tau_{2}$ horizon. In the limit $\tau_{2}\to0$
with $\tau_{1}$ fixed, with the correct expansions (\ref{smallk},\ref{largek}),
the propagator \eqref{eq:prop-basic} becomes

\[
F_{1L}^{\mathrm{I}}(x_{1},x_{2})\approx\frac{\lambda H^{2}}{(2\pi)^{2}}\frac{(\tau_{1}\tau_{2})^{\epsilon}}{3}\left[\frac{(\Lambda_{IR}\tau_{1})^{2\epsilon}}{(2\epsilon)^{2}}+\frac{1}{2}\left(\left(\frac{\mu}{H}\right)^{2}+\frac{1}{\epsilon}\right)\log\frac{\tau_{1}}{\tau_{2}}\right]\left(\cos1+\mathcal{O}(\epsilon)\right),
\]
which is well-behaved for $\tau_{2}\to0$.

\paragraph{The intermediate region:}

Region II consists of modes that are outside the current horizon,
but were inside the horizon at the time of the interaction, $\tau_{v}$.
Using the relevant expansions for the mode functions \eqref{mode}
we find the momentum space propagator

\begin{eqnarray*}
 &  & F_{1L}^{II}(k,\tau_{1},\tau_{2})\approx-\frac{\lambda H^{2}}{(2\pi)^{2}}\frac{(k^{2}\tau_{1}\tau_{2})^{\epsilon}}{4k^{4}}\int_{\tau_{0}}^{-k^{-1}}\frac{d\tau_{v}}{\tau_{v}^{2}}\sin(2k\tau_{v}-\pi\epsilon)\left[\frac{1-(\Lambda_{IR}\tau_{v})^{2\epsilon}}{2\epsilon}+\frac{1}{2}\left(\frac{\mu}{H}\right)^{2}\right]\\
 &  & \approx-\frac{\lambda H^{2}}{(2\pi)^{2}}\frac{(k^{2}\tau_{1}\tau_{2})^{\epsilon}}{4k^{3}}\left[\left(\frac{1}{2}\left(\frac{\mu}{H}\right)^{2}+\frac{1}{2\epsilon}\right)\int_{1}^{\infty}\frac{ds}{s^{2}}\sin(2s-\pi\epsilon)-\frac{\left(\frac{\Lambda_{IR}}{k}\right)^{2\epsilon}}{2\epsilon}\int_{1}^{\infty}\frac{ds}{s^{2-2\epsilon}}\sin(2s-\pi\epsilon)\right],
\end{eqnarray*}
where in the second line we changed variables $s\equiv-k\tau_{v},$
and as the main contribution to the integral comes from the lower
limit, we moved the upper limit to infinity. Written thus, the two
remaining integrals match to the leading order in $\epsilon$, yielding
$I\equiv\int_{1}^{\infty}\frac{ds}{s^{2}}\sin(2s-\pi\epsilon)\approx0.063+\mathcal{O}(\epsilon)$.
Converting back to position space we then have

\begin{align*}
F_{1L}^{II}(x_{1},x_{2}) & \approx-\frac{\lambda H^{2}I}{(2\pi)^{4}}\frac{(\tau_{1}\tau_{2})^{\epsilon}}{2}\int_{-\tau_{0}^{-1}}^{-\tau_{2}^{-1}}dk\left[\left(\frac{1}{2}\left(\frac{\mu}{H}\right)^{2}+\frac{1}{2\epsilon}\right)k^{-1+2\epsilon}-\frac{\Lambda_{IR}^{2\epsilon}}{2\epsilon}k^{-1}\right]\\
\approx & -\frac{\lambda H^{2}I}{8(2\pi)^{4}\epsilon}\left(\frac{\tau_{1}}{\tau_{2}}\right)^{\epsilon}\left[\left(\left(\frac{\mu}{H}\right)^{2}+\frac{1}{\epsilon}\right)\left(1-\left(\frac{\tau_{2}}{\tau_{0}}\right)^{2\epsilon}\right)-2(\Lambda_{IR}\tau_{0})^{2\epsilon}\left(\frac{\tau_{2}}{\tau_{0}}\right)^{2\epsilon}\log\frac{\tau_{0}}{\tau_{2}}\right].
\end{align*}

This result should be contrasted with the infrared contribution \eqref{eq:late-ir}.
Both asymptote to a constant value for very late times, but the IR
contribution is dominant due to an extra factor of $\epsilon^{-1}.$
One can again verify that the apparent divergence as $\tau_{2}\to0$
with $\tau_{1}$ fixed is due to the assumption $\tau_{1}\sim\tau_{2}$,
and vanishes as one computes the potentially divergent part more carefully.

\paragraph{The UV modes:}

Finally, in region III the scale $k$ is inside the current horizon
between $\tau_{0}$ and $\tau_{{\rm 2}}$. Using the expansion (\ref{largek})
the contribution to the propagator \eqref{k-space} becomes 
\begin{align}
F_{1L}^{III}(k,\tau_{1},\tau_{2}) & \approx-\frac{\lambda H^{2}}{(2\pi)^{2}}\frac{\tau_{1}\tau_{2}}{2k^{2}}\int_{\tau_{0}}^{\tau_{2}}\frac{d\tau_{v}}{\tau_{v}^{2}}\cos k(\tau_{1}-\tau_{v})\sin k(\tau_{2}-\tau_{v})\left[\frac{1-(\Lambda_{IR}\tau_{v})^{2\epsilon}}{2\epsilon}+\frac{1}{2}\left(\frac{\mu}{H}\right)^{2}\right]\nonumber \\
 & \approx-\frac{\lambda H^{2}}{(2\pi)^{2}}\frac{\tau_{1}\tau_{2}}{2k}\int_{\infty}^{-k\tau_{2}}\frac{ds}{s^{2}}\cos(s_{1}-s)\sin(s_{2}-s)\left[\frac{1-(\Lambda_{IR}/k)^{2\epsilon}s^{2\epsilon}}{2\epsilon}+\frac{1}{2}\left(\frac{\mu}{H}\right)^{2}\right],
\end{align}
where we again defined $s\equiv-k\tau_{v}$ and moved the upper limit
to infinity. To leading order in $\epsilon$ we then find 

\begin{align*}
F_{1L}^{III}(k,\tau_{1},\tau_{2}) & \approx-\frac{\lambda H^{2}}{(2\pi)^{2}}\frac{\tau_{1}\tau_{2}}{2k}\left[\frac{1-(\Lambda_{IR}\tau_{2})^{2\epsilon}}{2\epsilon}+\frac{1}{2}\left(\frac{\mu}{H}\right)^{2}\right]\\
 & \cdot\left(\cos k(\tau_{1}+\tau_{2})\mathrm{Ci}(-2k\tau_{2})-\sin k(\tau_{1}+\tau_{2})\left(\frac{\pi}{2}+\mathrm{Si}(2k\tau_{2})\right)\right),
\end{align*}
where the Cosine and Sine integrals are defined as

\[
\mathrm{Ci}(x)=-\int_{x}^{\infty}\frac{\cos s}{s}ds,\qquad\mathrm{and\qquad}\mathrm{Si}(x)=\int_{0}^{x}\frac{\sin s}{s}ds.
\]
Transforming back to position space we then get

\begin{align*}
F_{1L}^{III}(x_{1},x_{2}) & \approx-\frac{\lambda H^{2}}{(2\pi)^{4}}(\tau_{1}\tau_{2})\left[\frac{1-(\Lambda_{IR}\tau_{2})^{2\epsilon}}{2\epsilon}+\frac{1}{2}\left(\frac{\mu}{H}\right)^{2}\right]\\
 & \cdot\int_{-\tau_{2}^{-1}}^{\Lambda_{UV}a(\tau_{2})}dk\; k\left(\cos k(\tau_{1}+\tau_{2})\mathrm{Ci}(2k\tau_{2})-\sin k(\tau_{1}+\tau_{2})\left(\frac{\pi}{2}+\mathrm{Si}(2k\tau_{2})\right)\right)\\
 & \sim-\frac{\lambda H^{2}}{(2\pi)^{4}}\left(\frac{\tau_{1}}{\tau_{2}}\right)\left[\frac{1-(\Lambda_{IR}\tau_{2})^{2\epsilon}}{2\epsilon}+\frac{1}{2}\left(\frac{\mu}{H}\right)^{2}\right].
\end{align*}
Above, we noted that the upper limit of the integral doesn't contribute
due to the oscillatory nature of the integrand, and then obtained
a rough estimate of the magnitude of the integral by the value of
the integrand at the lower limit multiplied by frequency of oscillation.
We again find a contribution that asymptotes to a constant for very
late times, but is sub-dominant to the infrared and intermediate contributions
found above.

\subsection{The one-loop propagators \label{sub:full-prop}}

We can now write down the propagators to one loop using the results
above. The tree level contributions from the infrared modes \eqref{smallk}
to the propagators can be written as

\begin{align}
F_{0L}(x_{1},x_{2}) & =\int\frac{d^{3}\vec{k}}{(2\pi)^{3}}e^{i\vec{k}\cdot(\vec{x}_{1}-\vec{x}_{2})}F(k,\tau_{1},\tau_{2})\approx\frac{H^{2}}{(2\pi)^{2}}(\tau_{1}\tau_{2})^{\epsilon}\int_{\Lambda_{IR}}^{\min(-\tau_{1}^{-1},-\tau_{2}^{-1})}dk\: k^{-1+2\epsilon}\label{eq:F-tree}\\
 & \approx\frac{H^{2}}{(2\pi)^{2}}\left(\frac{\tau_{1}\tau_{2}}{\min(\tau_{1},\tau_{2})^{2}}\right)^{\epsilon}\frac{1-(\Lambda_{IR}\min(\tau_{1},\tau_{2}))^{2\epsilon}}{2\epsilon},\nonumber \\
G^{R}(x_{1,}x_{2}) & \approx\frac{2H^{2}}{3(2\pi)^{2}}\theta(\tau_{1}-\tau_{2})\left(\tau_{1}^{3-\epsilon}\tau_{2}^{\epsilon}-\tau_{1}^{\epsilon}\tau_{2}^{3-\epsilon}\right)\int_{\Lambda_{IR}}^{-\tau_{1}^{-1}}dk\: k^{2}\nonumber \\
 & \approx-\frac{2H^{2}}{9(2\pi)^{2}}\theta(\tau_{1}-\tau_{2})\left(\frac{\tau_{2}}{\tau_{1}}\right)^{\epsilon}\left(1-\left(\frac{\tau_{2}}{\tau_{1}}\right)^{3-2\epsilon}\right).\label{eq:Gfull}
\end{align}
Adding the dominant one loop IR contribution \eqref{eq:late-ir} to
the tree contribution $F_{0L}(x_{1},x_{2})$, we find the full propagator
for late times as

\begin{equation}
F(x_{1},x_{2})\approx\frac{H^{2}}{(2\pi)^{2}}\frac{1}{2\epsilon}\left(1-\frac{\lambda}{12(2\pi)^{2}\epsilon}\left[\left(\frac{\mu}{H}\right)^{2}+\frac{1}{\epsilon}\right]\left(\left(\frac{\tau_{1}}{\tau_{2}}\right)^{\epsilon}+\left(\tau_{1}\leftrightarrow\tau_{2}\right)\right)\right),\label{eq:Ffull}
\end{equation}
where the symmetrization $\tau_{1}\leftrightarrow\tau_{2}$ corresponds
to including the mirror images of the diagrams in figure \ref{fig-loop}.
This shows that for a massive field (where $\epsilon>0$) the comoving
infrared cutoff $\Lambda_{IR}$ drops out of the late time correlators,
which match those obtained from a Bunch-Davies/Euclidean vacuum computation.

\subsection{Early time propagator \label{sub:Early}}

We should contrast the propagator found above with the corresponding
result for early times or very light fields. The infrared contribution
to the tree level propagator for light (or massless) fields can be
computed from \eqref{eq:F-tree} by a power series expansion in $\epsilon$
as 

\begin{align}
F_{0L}^{\mathrm{early}}(x_{1},x_{2}) & \approx\frac{H^{2}}{(2\pi)^{2}}\left(\log(\Lambda_{IR}\tau_{0})+\log\left(\frac{\min(\tau_{1},\tau_{2})}{\tau_{0}}\right)\right)\left(1+\mathcal{O}(\epsilon)\right),\label{eq:early1}
\end{align}
where the first logarithm measures how far the infrared cutoff is
from the initial horizon, and the second logarithm gives the number
of e-folds of inflation. The one-loop correction can be similarly
computed by expanding \eqref{eq:Fir} as a power series in $\epsilon.$
After some algebra this gives

\begin{align}
F_{1L}^{I}(x_{1},x_{2}) & \approx\frac{\lambda H^{2}}{9(2\pi)^{4}}\left[\log^{3}\frac{\tau_{2}}{\tau_{0}}+3\log^{2}\frac{\tau_{2}}{\tau_{0}}\left(\log(\Lambda_{IR}\tau_{0})-\frac{1}{8}\left(\frac{\mu}{H}\right)^{2}\right)\right.\nonumber \\
 & -\frac{1}{4}\log\frac{\tau_{2}}{\tau_{1}}\left(\left(\frac{\mu}{H}\right)^{2}+\log(\Lambda_{IR}\tau_{0})\left(\frac{3}{4}\left(\frac{\mu}{H}\right)^{2}-3\log(\Lambda_{IR}\tau_{0})\right)\right)\nonumber \\
 & +\left.\log(\Lambda_{IR}\tau_{0})\left(\log(\Lambda_{IR}\tau_{0})-\frac{1}{8}\left(\frac{\mu}{H}\right)^{2}\right)-\frac{1}{12}\left(\frac{\mu}{H}\right)^{2}\right]\left(1+\mathcal{O}(\epsilon)\right).\label{eq:early}
\end{align}
These results showcase the familiar logarithmic divergences for late
times $\tau_{2}\to0$, but as shown earlier, these are absent for
truly late times if $\epsilon>0$. The solution to this apparent paradox
is that the expansions in \eqref{eq:early1}, \eqref{eq:early} are
valid when $\epsilon\ll1/\log\left(\tau_{0}/\tau_{2}\right),$ which
means the logarithmic expansion breaks down for sufficiently late
times, and the propagator is better described by \eqref{eq:Ffull},
which is well behaved.

\section{DISCUSSION AND CONCLUSIONS }

\paragraph{Massive scalar }

We conclude from section \ref{sub:full-prop} that in the distant
future, the loop corrections are insensitive to the choice of infrared
cutoff. The result for the comoving cutoff approaches that of the
de Sitter invariant Euclidean vacuum results as $\tau_{2}^{2\epsilon}$
, i.e. as a negative power of the scale factor.

\paragraph{Massless scalar, or a light scalar at early times}

Here the result of section \ref{sub:Early} is applicable. This can
be the relevant behavior during the entire course of slow roll inflation.
The difference between the result of section \ref{sub:Early} and
section \ref{sub:full-prop} provides a measure of the theoretical
uncertainty in the predictions of the two-point function caused by
uncertainties in the initial state at the start of inflation. This
difference is proportion to a $\lambda\log^{2}\left(\tau_{2}/\tau_{0}\right)\log\left(\Lambda_{IR}\tau_{0}\right)$
relative to the tree level contribution. The slow roll conditions
for pure $\lambda\phi^{4}$ give the constraint that the initial expectation
value for $\phi$ must be larger than the Planck scale. The condition
that the density fluctuations today be sufficiently small $(10^{-5})$
gives the constraint that $\lambda\approx10^{-15}$. So we see in
this example the effect will be at least $10^{-11}$ smaller than
the tree-level piece, even after 60 e-folds of inflation. We conclude
therefore, that while these secular logarithms are present during
primordial slow-roll inflation, matching the magnitude of our observed
density perturbations constrains the parameters of the inflaton model
such that the coefficient of the secular log is very small, unless
inflation is very long-lived.

\paragraph{Massless scalar at late times}

The Euclidean vacuum methods break down in this case (see for example
\cite{Hollands:2011we} for a recent discussion of this situation)
due to the nonexistence of a de Sitter invariant vacuum for a massless
scalar \cite{Allen:1985ux}. However the comoving IR cutoff approach
still provides a well-defined perturbative expansion, up until the
point when $\lambda\log^{3}\left(\tau_{2}/\tau_{0}\right)\sim1$ (see
\eqref{eq:early}). The prediction is that the theoretical uncertainty
in correlators becomes large at sufficiently late times. This is the
practical sense in which de Sitter spacetime suffers a quantum instability.
In the slow roll example, this does not happen until about $10^{5}$
e-folds, so is really only of academic interest as far as primordial
inflation goes.

\paragraph{Graviton at late times}

However it should be noted our result for the massless scalar matches
qualitatively with the computation of \cite{Tsamis:1996qk}. There
the full one-loop correction for graviton self-energy was computed
with the same kind of comoving infrared cutoff that we have used in
the present paper. Compared to the tree-level result the contribution
is of order $\left(\frac{H}{M_{pl}}\right)^{4}\log^{2}\left(\tau_{2}/\tau_{0}\right)$.
Again this is very small over the 60 e-folds needed for primordial
inflation. However over the present epoch such a term may become important,
if we are indeed evolving toward a period where dark energy dominates.
To apply these ideas to our present epoch, we could imagine simply
shifting $\tau_{0}$ to the present time, and placing a comoving infrared
cutoff just outside our present horizon. This suggests quantum uncertainties
due to the indeterminancy of our initial state would only become large
after $\left(\frac{M_{pl}}{H_{0}}\right)^{2}\approx10^{122}$ e-folds
of late time expansion, where now $H_{0}\approx10^{-33}$ eV is the
Hubble parameter today. It is interesting that this provides a mechanism
for the possible breakdown of semiclassical physics in a de Sitter
region, as predicted by quantum gravity arguments in \cite{Lowe:2010np}.
\begin{acknowledgments}
This research is supported in part by DOE grant DE-FG02-91ER40688-Task
A and an FxQI grant. We thank Richard Easther, Richard Woodard and
Helmut Friedrich for helpful discussions.
\end{acknowledgments}
\bibliographystyle{kp}
\bibliography{desittir}

\end{document}